\documentclass[journal]{IEEEtran}
\usepackage{amssymb,url}
\usepackage{graphicx,epsfig,color,wrapfig}

\title{On R\'enyi and Tsallis entropies and divergences for exponential families}

\author{Frank~Nielsen,~\IEEEmembership{Senior Member,~IEEE}  and    Richard Nock  ~\IEEEmembership{Nonmember,~IEEE}  
\thanks{F. Nielsen is with the Department
of Fundamental Research of Sony Computer Science Laboratories, Inc., Tokyo, Japan, and the Computer Science Department (LIX) of \'Ecole Polytechnique, Palaiseau, France. 
 e-mail: Frank.Nielsen@acm.org}
\thanks{R. Nock is with the Economics Department (CEREGMIA) of the University of Antilles-Guyane, France. e-mail: rnock@martinique.univ-ag.fr}
\thanks{Manuscript received May 2011}
}

\newenvironment{proof}{\vspace{1ex}\noindent{\bf Proof}\hspace{0.5em}}
	{\hfill\qed\vspace{1ex}}
\newenvironment{example}{\vspace{1ex}\noindent{\bf Example}\hspace{0.5em}}
	{\hfill$\Diamond$\vspace{1ex}}
	
\def\dx{\mathrm{d}x}
\def\Inner#1#2{ {\left\langle #1, #2 \right\rangle} }
\def\inner#1#2{ {\langle #1, #2 \rangle} }
\def\innerproduct#1#2{ {\langle #1, #2 \rangle} }
\def\Innerproduct#1#2{ {\left\langle #1, #2 \right\rangle} }
\def\dotproduct#1#2{ {\langle #1, #2 \rangle} }
\def\KL{\mathrm{KL}}
\def\d{\mathrm{d}}
\def\dalpha{\mathrm{d}\alpha}

\def\framethis#1{#1}

\begin{document}
\maketitle

\begin{abstract}
Many common probability distributions in statistics like the Gaussian, multinomial, Beta or Gamma distributions can be studied under the unified framework of
exponential families.
In this paper, we prove that both R\'enyi and Tsallis divergences of distributions belonging to the same exponential family admit a generic closed form expression. 
Furthermore, we show that R\'enyi and Tsallis entropies can also be calculated in closed-form for sub-families including the Gaussian or exponential distributions, among others.   
\end{abstract}


\IEEEpeerreviewmaketitle

%%%%%%%%%%%%%%%%%%%%%%
\section{Introduction}
%%%%%%%%%%%%%%%%%%%%%%

In 1948, Shannon published a theory on communications that initiated the field of information theory~\cite{Shannon:1948}.
Nowadays, it is well-known that Shannon {\it entropy} quantitatively measures the amount of {\it uncertainty}~\cite{ct-1991} of a random variable.
The entropy $H(P)$ of a random variable $P$ is defined according to its underlying density $p(x)$ as

\begin{eqnarray}
H(P) &=& \int p(x)\log \frac{1}{p(x)}\dx\\
& =& - \int p(x)\log p(x) \dx = E_P[-\log p(x)]. 
\end{eqnarray}

Closed-form expressions for the Shannon entropy for many continuous distributions are reported in~\cite{Entropy:2000}.
In coding theory, one seeks for  codes that uses the underlying structure of the message language at its best.
Since in practice, the true model distribution $P$ is hidden by nature and therefore unknown to the observer,  
 we rather define the {\it cross-entropy} between the considered model $Q$ and the unknown ideal random variables $P$ as 
 
\begin{eqnarray}
H^\times(P : Q) &=& E_P[-\log q(x)] = -\int p(x)\log q(x) \dx\\
&\geq&  H^\times(P : P).
\end{eqnarray}

It follows that the Kullback-Leibler divergence  (also called relative entropy) between two distributions $P$ and $Q$  is defined  by

\begin{equation}
\KL( P : Q ) = \int_x p(x)\log\frac{p(x)}{q(x)} \dx=E_P \left[\log\frac{p(x)}{q(x)}\right]
\end{equation}

The Kullback-Leibler divergence  $\KL(P:Q)$ is an oriented distance (i.e., $\KL(P:Q)\not = \KL(Q:P)$, emphasized by the ``:'' notational convention) that can be rewritten as

\begin{equation}
\KL(P:Q) =  H^\times(P:Q)- H(P) \geq 0
\end{equation}

In 1961, R\'enyi generalized the Shannon entropy by modifying one of its axiom characterizing the {\it averaging} of information.
The R\'enyi $H_\alpha(p)$ entropy~\cite{Renyi-1961} of a probability distribution $p$ is a single-parametric function  defined by

\begin{equation}
H^R_\alpha(p) =   \frac{\log \int p^\alpha(x) \dx}{1-\alpha}, \alpha\in (0,+\infty)\backslash\{1\}.
\end{equation}

Let us prove using L'H\^opital rule\footnote{L'H\^opital rule dates back to the 17th century, and states that the limit of the indeterminate ratio of functions equals to the limit of the ratio of their derivatives provided that 
(i) the limits of both the numerator and denominator coincide, and that
(ii) the limit of the ratio of the derivatives also exists.
That is, if $\lim_{x \to \alpha} f(x) = \lim_{x \to \alpha} g(x)=0$ and  $\lim_{x\to \alpha} f'(x)/g'(x)=l$  exists,
then 
$\lim_{x\to \alpha}\frac{f(x)}{g(x)} = \lim_{x\to \alpha}\frac{f'(x)}{g'(x)}=l$.} that R\'enyi entropies tend to Shannon entropy $H(p)=-\int p(x)\log p(x)\dx$ when $\alpha\rightarrow 1$.
(This is a classical proof explained in textbooks that we include here to illustrate L'H\^opital rule that we shall use repeatedly later on.)

\begin{proof}
Consider the discrete case (i.e., counting measure) of R\'enyi and Shannon entropies.
Set $f(\alpha)=\log \sum_{i=1}^n p_i^\alpha$ (for any fixed distribution $P$) and $g(\alpha)=1-\alpha$. 
Then $\frac{\d g(\alpha)}{\dalpha}=-1$ and 
\begin{equation}
\frac{\d f(\alpha)}{\dalpha} = \frac{\sum_{i=1}^n \frac{\d}{\dalpha} (p_i^\alpha) }{\sum_{i=1}^n p_i^\alpha}
\end{equation}
after applying the derivative chain rule.
Since 
\begin{equation}
\frac{\d}{\dalpha} (p_i^\alpha) = \frac{\d}{\dalpha} e^{\alpha\log p_i} = (\log p_i) e^{\alpha\log p_i} = p_i^\alpha \log p_i,
\end{equation}
 we get 
\begin{equation}
\frac{f'(\alpha)}{g'(\alpha)}= -  \sum_{i=1}^n p_i^\alpha \log p_i, \mbox{\ and\ }  \lim_{\alpha \to 1 } \frac{f'(\alpha)}{g'(\alpha)}= -  \sum_{i=1}^n p_i \log p_i.
\end{equation}
Since $\lim_{\alpha\to 1} f(\alpha)=\lim_{\alpha\to 1} g(\alpha)=0$ and $\lim_{\alpha\to 1} \frac{f'(\alpha)}{g'(\alpha)}= -  \sum_{i=1}^n p_i \log p_i$, we deduce from l'H\^opital rule that $\lim_{\alpha\to 1} H^R_{\alpha}(P)=H(P)$.
That is, R\'enyi entropy tends to Shannon entropy as $\alpha\to 1$.
\end{proof}

R\'enyi entropies keep Shannon additivity property~\cite{ct-1991} of independent systems, and are concave and monotonically decreasing function of $\alpha$.
Closed-form formula for the R\'enyi entropies of many multivariate distributions are reported in~\cite{RenyiClosedForm:2005}, and for the multivariate Gaussian distribution in the technical report~\cite{JensenRenyi-CSPL328-2001}. 

In 1988, Tsallis (motivated by physical multi-fractal systems) introduced yet another one-parameter generalization of Shannon entropy.
Historically, this family of entropic functions was derived axiomatically by Havrda and Charvat~\cite{Havrda-Charvat:67} in 1967.
Tsallis $H^T_\alpha(p)$ entropies of a probability distribution $p$ are defined by

\begin{equation}
H^T_\alpha(p) =   \frac{\int p(x)^\alpha\dx -1}{1-\alpha}, \alpha\in \mathbb{R}\backslash\{1\}
\end{equation}
Tsallis entropies are non-additive, tending to Shannon entropy when $\alpha\rightarrow 1$, and can be derived from the generalized Shannon-Khinchin axioms~\cite{Tsallis2Div:2009}.

Let $I_\alpha(p)=\int p(x)^\alpha \dx$, then R\'enyi and Tsallis entropies can be rewritten as

\begin{eqnarray}
H^R_\alpha(p) &=&   \frac{\log I_\alpha(p)}{1-\alpha}, \label{eq:IR} \\
H^T_\alpha(p) & = &  \frac{I_\alpha(p) -1}{1-\alpha}.  \label{eq:IT}
\end{eqnarray}

Since $I_\alpha(p)=(1-\alpha)H^T_\alpha(p) +1=e^{(1-\alpha)H^R_\alpha(p)}$ , we can convert these two families of entropies through 
the following monotonic conversion functions:

\begin{eqnarray}
H^R_\alpha(p) &=& \frac{\log ((1-\alpha) H^T_\alpha(p) +1) }{1-\alpha}\\ 
H^T_\alpha(p) & =&  \frac{e^{(1-\alpha)H^R_\alpha(p)}-1}{1-\alpha}  \label{eq:RT}
\end{eqnarray}

%%%%%%%%%%%%%%%%%%%%%%%%%%%%%%%%%%%%%%%%%%%%%%%%%%%%%%%%
\section{R\'enyi and Tsallis entropies of exponential families}\label{sec:ent}
%%%%%%%%%%%%%%%%%%%%%%%%%%%%%%%%%%%%%%%%%%%%%%%%%%%%%%%%

A random variable $X\sim E_F(\theta)$ is said to belong to the exponential family $E_F$~\cite{EF:1978} when it admits the following canonical decomposition of its density:

\begin{equation}
p_F(x;\theta)=\exp \left( \innerproduct{t(x)}{\theta}-F(\theta)+k(x) \right),
\end{equation}

where $\innerproduct{x}{y}=x^T y$ denotes the inner product, $t(x)$ the sufficient statistics, $\theta$ the natural parameters, $F(\theta)$ a $C^\infty$ differentiable real-valued convex function, and $k(x)$ a carrier measure.

Since $F(\theta)=\log \int_x \exp (\innerproduct{t(x)}{\theta}+k(x)) \dx$ (because $\int p_F(x;\theta)\dx=1$), function $F$ is called the log-normalizer.
Function $F$ characterizes the family, while the natural parameter $\theta$ denotes the member of the family $E_F$.

A statistic is a function of the observations (say, the sample mean or sample variance) that collects information about the distribution with the goal to concentrate information for later inference.
A statistic is said sufficient if it allows one to concentrate information obtained from random observations {\em without loosing} information, in a sense that working directly on the observation sets or its compact sufficient statistics yields exactly the same parameter estimation results.
It can be shown from the Neyman-Pearson factorization theorem~\cite{ct-1991}, under mild regularity conditions, that the class of distributions admitting sufficient statistics are  precisely the exponential families~\cite{ef-flashcards-2009}.
Term $k(x)$ is related to the carrier measure (i.e., counting or Lebesgue).
An exponential family can be univariate (eg., like the Poisson or 1D Gaussian distributions) or multivariate (like the multinomial or $d$-dimensional Gaussian distributions). The order of an exponential family denotes the dimension of the parameter space.
Thus the Gaussian (normal) distribution is univariate of order $2$ (parameters $\mu$ and $\sigma$).

Many common distribution families such as Poisson, Gaussian or multinomial distributions are exponential families whose canonical decompositions 
$(F,t,\theta,k)$ are given in~\cite{ef-flashcards-2009}.

Let us prove that for any distribution belonging to the exponential families, we have the following entropy expressions:

\begin{eqnarray}
\lefteqn{H^R_\alpha(p_F(x;\theta))} \nonumber\\ &=&  \frac{1}{1-\alpha} \left(F(\alpha\theta)-\alpha F(\theta)+\log E_p[e^{(\alpha-1)k(x)}]\right)  
\end{eqnarray}

\begin{eqnarray}
\lefteqn{H^T_\alpha(p_F(x;\theta))} \nonumber\\ &=&   \frac{1}{1-\alpha} \left((e^{F(\alpha\theta)-\alpha F(\theta)}) E_p[e^{(\alpha-1)k(x)}]-1  \right)
\end{eqnarray}

\begin{proof}

Consider calculating  $I_\alpha(p)=\int p(x)^\alpha \dx$ term for exponential families:

\begin{eqnarray}
&&I_\alpha(p) = \int e^{\alpha (\inner{t(x)}{\theta}-F(\theta)+k(x)) } \dx  \\
& = & \int e^{\inner{t(x)}{\alpha\theta}-\alpha F(\theta)+\alpha k(x) + (1-\alpha) k(x) - (1-\alpha) k(x) + F(\alpha\theta)-F(\alpha\theta)} \dx \\
& = & \int e^{F(\alpha\theta)-\alpha F(\theta)} p_F(x;\alpha\theta) e^{(\alpha-1) k(x)} \dx\\
& = & e^{F(\alpha\theta)-\alpha F(\theta)}  \int p_F(x;\alpha\theta) e^{(\alpha-1) k(x)} \dx \\
& =& e^{F(\alpha\theta)-\alpha F(\theta)}  E_p[e^{(\alpha-1) k(x)}]
\end{eqnarray}

The formula for the R\'enyi and Tsallis entropies are then derived using Eq.~\ref{eq:IR} and Eq.~\ref{eq:IT}.

\end{proof}

In particular, for  standard carrier measure $k(x)=0$ (eg., Gaussian, exponential, Bernoulli or centered Laplacian), we obtain the following  generic {\it closed-form} expressions of R\'enyi and Tsallis entropies:

\begin{eqnarray}
H^R_\alpha(p_F(x;\theta)) &=&  \frac{1}{1-\alpha} \left(F(\alpha\theta)-\alpha F(\theta) \right) \label{eq:RE}  \\
H^T_\alpha(p_F(x;\theta)) &=&   \frac{1}{1-\alpha} \left(e^{F(\alpha\theta)-\alpha F(\theta)}-1 \right) \label{eq:TE}
\end{eqnarray}

For $\alpha\rightarrow 1$, observe that both those formula  yields  to Shannon entropy for exponential families (with $k(x)=0$):

\begin{equation} \label{eq:SE}
H(p_F(x;\theta)) = F(\theta) - \inner{\theta}{\nabla F(\theta)} 
\end{equation}

\begin{proof}
Let us use L'H\^opital rule on R\'enyi entropy

\begin{eqnarray}
H_\alpha(p_F(x;\theta)) &=&  \frac{1}{1-\alpha} \left(F(\alpha\theta)-\alpha F(\theta) \right) \\
& \simeq_{\alpha\rightarrow 1} & \frac{\inner{\theta}{\nabla F(\alpha\theta)} -F(\theta) }{-1} \\
& \simeq_{\alpha\rightarrow 1} & F(\theta)-\inner{\theta}{\nabla F(\theta)}
\end{eqnarray}
(using G\^ateaux derivatives $\nabla_\alpha F(\alpha\theta)=\inner{\theta}{\nabla F(\alpha\theta)}$)
\end{proof}

For non-zero carrier measure the Shannon entropy of an exponential family $p\sim E_F(\theta)$ is
$H(p_F(x;\theta)) = F(\theta) - \inner{\theta}{\nabla F(\theta)} - E_p[k(x)]$. This will be proved in section~\ref{sec:div}.

%Table~\ref{tab:cd} describes the canonical decompositions of common exponential family distribution admitting standard carrier measure $k(x)=0$.
%\begin{table}
%\caption{Some canonical decompositions of exponential families (with $k(x)=0$).}\label{tab:cd}
%\end{table}

\begin{example}
To illustrate the generic entropy formula, let us start with a simple exponential family: the exponential distribution.
The exponential distribution models the time between two successive Poisson processes, and has  density

\begin{equation}
p(x;\lambda) = \lambda e^{-\lambda x}, x\geq 0
\end{equation}
where $\lambda>0$ is called the rate parameter.

Writing $\lambda e^{-\lambda x}=e^{-\lambda x+\log \lambda}$, we get the canonical decomposition of exponential families with 
$t(x)=x$, $\theta=-\lambda$, $F(\theta)=-\log\lambda=-\log -\theta$ and $k(x)=0$.
The exponential distribution is a univariate exponential family of order $1$.
The R\'enyi entropy is $H^R_\alpha(p)=\frac{1}{1-\alpha}(F(\alpha\theta)-\alpha F(\theta))=\frac{1}{1-\alpha}(-\log \alpha\lambda+\alpha\log\lambda)=\log\lambda-\frac{\log\alpha}{1-\alpha}$.
The log-normalizer derivative is $F'(\theta)=-\frac{1}{\theta}=\frac{1}{\lambda}$.
The Shannon entropy is $H(p)=F(\theta)-\theta F'(\theta)=1-\log \lambda$.
Using L'H\^opital rule, we find that $\lim_{\alpha\rightarrow 1} H^R_\alpha(p)=-\log\lambda-\lim_{\alpha\rightarrow 1} \frac{\log\alpha}{1-\alpha}=1-\log\lambda = H(P)$.
Tsallis entropy is $\frac{1}{1-\alpha} (\frac{\lambda^\alpha}{\alpha\lambda}-1)=\frac{\lambda^\alpha-\alpha\lambda}{\alpha(1-\alpha)\lambda}$.
Again, using L'H\^opital rule, we find that Tsallis entropy converges to Shannon entropy as $\alpha\rightarrow 1$:
$H^T_\alpha(p)\lim_{\alpha\rightarrow 1} \frac{(\lambda^\alpha-\alpha\lambda)'}{(\alpha(1-\alpha)\lambda)'}=\frac{\lambda\log\lambda-\lambda}{-\lambda}=1-\log\lambda=H(p)$ (where the derivatives are computed according to parameter $\alpha$).
\end{example}

\begin{example}
Let us consider now the usual Gaussian distribution (univariate of order $2$) with density 
\begin{equation}
p(x;\mu,\sigma)=\frac{1}{\sqrt{2\pi\sigma^2}}  e^{-\frac{(x-\mu)^2}{2\sigma^2}}.
\end{equation}

Its canonical decomposition into an exponential family yields 

\begin{eqnarray}
&& p(x;\mu,\sigma) = \frac{1}{\sqrt{2\pi\sigma^2}}  e^{-\frac{(x-\mu)^2}{2\sigma^2}} \\
& = & \exp \left( -\frac{x^2}{2\sigma^2}+x\frac{\mu}{\sigma^2}-\frac{\mu^2}{2\sigma^2}-\frac{1}{2}\log 2\pi\sigma^2\right) \\
&=& \exp \left(  \Inner{(x,x^2)}{\left(\frac{\mu}{\sigma^2},-\frac{1}{2\sigma^2}\right)}-F(\theta)  \right)
\end{eqnarray}

\begin{itemize}
\item Sufficient statistics: $t(x)=(x,x^2)$, 
\item Natural parameters $\theta=(\frac{\mu}{\sigma^2},-\frac{1}{2\sigma^2})$,
\item Log-normalizer 
$F(\theta)= -\frac{\theta_1^2}{4\theta_2}+\frac{1}{2}\log\frac{2\pi}{-\theta_2}  =\frac{\mu^2}{2\sigma^2}+\frac{1}{2}\log 2\pi\sigma^2$, and 
\item Carrier measure $k(x)=0$.
\end{itemize}

Thus the R\'enyi entropy of Eq.~\ref{eq:RE} instanced to the Gaussian case is 
 
\begin{eqnarray}
H^R_\alpha(p) &=&  \frac{1}{1-\alpha} \left(F(\alpha\theta)-\alpha F(\theta)\right) \\ 
& = & \frac{1}{1-\alpha} \left( \alpha\frac{\mu^2}{2\sigma^2}+\frac{1}{2}\log \frac{2\pi\sigma^2}{\alpha}-\alpha \left(\frac{\mu^2}{2\sigma^2}+\frac{1}{2}\log 2\pi\sigma^2\right)  \right) \\
& = & \frac{1}{1-\alpha} \left(
\frac{1-\alpha}{2}\log 2\pi\sigma^2-\frac{1}{2}\log\alpha
\right)\\& = & \frac{1}{2}\log 2\pi\sigma^2-\frac{\log\alpha}{2(1-\alpha)}
\end{eqnarray}

When $\alpha\rightarrow 1$, $H^R_\alpha(p) \rightarrow \frac{1}{2}\log 2\pi e\sigma^2$
(using L'H\^opital rule $\frac{\log\alpha}{2(\alpha-1)} \simeq_{\alpha\rightarrow 1} -\frac{1}{2}$).
Now using the Shannon closed form entropy of Eq.~\ref{eq:SE} with $\nabla F(\theta)=(-\frac{\theta_1}{2\theta_2},\frac{\theta_1^2}{4\theta_2^2}-\frac{1}{2\theta_2}) =(\mu,\mu^2+\sigma^2)$
, we again find the Gaussian entropy $H(p)=\frac{1}{2}\log 2\pi e\sigma^2$:

\begin{eqnarray}
H(p) &=& F(\theta)-\inner{\theta}{\nabla F(\theta)}\\
& = & \frac{\mu^2}{2\sigma^2}+\frac{1}{2}\log 2\pi\sigma^2-\Inner{\left(\frac{\mu}{\sigma^2},-\frac{1}{2\sigma^2}\right)}{(\mu,\mu^2+\sigma^2)}\\
&= & \frac{\mu^2}{2\sigma^2}+\frac{1}{2}\log 2\pi\sigma^2 - \frac{\mu^2}{\sigma^2}+\frac{\mu^2}{2\sigma^2}+\frac{1}{2}\\
&=& \frac{1}{2}\log 2\pi e\sigma^2
\end{eqnarray}

It follows from the conversion formula of Eq.~\ref{eq:RT} that the Tsallis entropy of the Gaussian is

\begin{equation}
H^T_\alpha(p) = \frac{e^{(1-\alpha)\log (2\pi e\sigma^2)^{\frac{1}{2}}}-1}{1-\alpha}=\frac{(2\pi e\sigma^2)^{\frac{1-\alpha}{2}} }{1-\alpha}.
\end{equation}

Again, we check that when $\alpha\rightarrow 1$, the Tsallis entropy tends to Shannon entropy:

\begin{eqnarray}
\frac{e^{(1-\alpha)\log (2\pi e\sigma^2)^{\frac{1}{2}}}-1}{1-\alpha} &\simeq_{\alpha\rightarrow 1}&  
\frac{1+(1-\alpha)\log (2\pi e\sigma^2)^{\frac{1}{2}} -1}{1-\alpha}\\
& = &  \frac{1}{2}\log 2\pi e\sigma^2
\end{eqnarray}

\end{example}

Similarly but with greater matrix calculus complexity, based on the canonical decomposition reported in~\cite{ef-flashcards-2009}, we may
consider the multivariate Gaussian distribution $X\sim \mathcal{N}(\mu,\Sigma)$ with mean $\mu$ and covariance matrix $\Sigma$  ($\det \Sigma=|\Sigma|>0$).
Appendix~\ref{sec:mg} provides the calculus details. We report the result here.
The R\'enyi $\alpha$-entropy is  given by

\begin{equation}
H^R_\alpha(X) = \frac{d}{2}\log 2\pi  +\frac{1}{2}\log |\Sigma| + \frac{d\log\alpha}{2(\alpha-1)}
\end{equation}
and tend to Shannon entropy as $\alpha\rightarrow 1$  (using L'H\^opital rule $\frac{d\log\alpha}{2(\alpha-1)} \simeq_{\alpha\rightarrow 1} \frac{d}{2}$):

\begin{equation}
H(X) = \frac{1}{2} \log (2\pi e)^d |\Sigma|
\end{equation}

Note that the sufficient statistics $t(x)$ does not intervene in the entropy formula.
The sufficient statistics plays a role for estimating parameter $\theta$ from independent and identically distributed (i.i.d.) observations, as mentioned in the concluding remarks (see Section~\ref{sec:concl}).

%%%%%%%%%%%%%%%%%%%%%%%%%%%%%%%%%%%%%%%%%%%%%%%%%%%%%%
\section{R\'enyi and Tsallis divergences of exponential families}\label{sec:div}
%%%%%%%%%%%%%%%%%%%%%%%%%%%%%%%%%%%%%%%%%%%%%%%%%%%%%%%%

Consider now two probability distributions $P$ and $Q$, and define the R\'enyi $D^R_\alpha(p:q)$ and Tsallis $D^T_\alpha(p:q)$ divergences as follows

%\framethis{
\begin{eqnarray}
D^R_\alpha(p:q) & = &  \frac{\log \int p(x)^\alpha q(x)^{1-\alpha} \dx }{\alpha-1} \\
D^T_\alpha(p:q) & = &  \frac{\int p(x)^\alpha q(x)^{1-\alpha} \dx - 1}{\alpha-1}  
\end{eqnarray}
%}

Those divergences are related to the $\alpha$-divergence\footnote{Historically, this divergence was first presented by Chernoff in~\cite{Chernoff-1952}.} 
\begin{equation}
I_{\alpha}(p:q)=\int p(x)^\alpha q(x)^{1-\alpha} \dx
\end{equation}
 that plays an important role\footnote{Namely, the role of canonical divergence in constant curvature statistical manifolds.} in
information geometry~\cite{informationgeometry-2000}:

\begin{eqnarray}
D^R_\alpha(p:q) & = &  \frac{\log I_{\alpha}(p:q) }{\alpha-1} \\
D^T_\alpha(p:q) & = &  \frac{I_{\alpha}(p:q) - 1}{\alpha-1}  
\end{eqnarray}

R\'enyi divergence can also be rewritten as 
\begin{equation}
D^R_\alpha(p:q)  =   \frac{\log \int p(x)^\alpha q(x)^{1-\alpha} \dx }{\alpha-1} =
\frac{\log \int \left(\frac{p(x)}{q(x)}\right)^\alpha q(x)\dx }{\alpha-1},
\end{equation}
that shows it is a Csisz\'ar $f$-divergence~\cite{Csiszar-1967}.
The special case $\alpha=\frac{1}{2}$ yields 
\begin{equation}
D^R_{\frac{1}{2}}(p:q)  = -2 \log \int \sqrt{p(x)} \sqrt{q(x)} \dx=-2\log B(p,q),
\end{equation}
where
$B(p,q)= \int \sqrt{p(x)} \sqrt{q(x)} \dx$ is called the Bhattacharrya coefficient~\cite{Bhatta1943}.
The Bhattacharrya coefficient is itself related to the (squared) Hellinger distance~\cite{hellinger-1907}:

\begin{eqnarray}
&& H^2(p:q)= \frac{1}{2} \int ( \sqrt{p(x)} - \sqrt{q(x)} )^2 \dx\\
& = & \frac{1}{2} \left( \int p(x)\dx + \int q(x)\dx -2\int (\sqrt{p(x)} \sqrt{q(x)}\dx \right)  \\
&=& 1 - B(p,q).
\end{eqnarray}

For members of the same exponential families (we do not require standard carrier measure $k(x)$ to be zero anymore), the  R\'enyi and Tsallis divergences~\cite{renyidivergence-2010} can always be calculated from the following closed-form solution:

\framethis{
\begin{eqnarray}
D^R_\alpha( p_F(x;\theta) : p_F(x;\theta') ) & = &   \frac{1}{1-\alpha} J_{F,\alpha}(\theta : \theta') \\
D^T_\alpha(( p_F(x;\theta) : p_F(x;\theta') ) & = &   \frac{1}{1-\alpha} \left( e^{-J_{F,\alpha}(\theta : \theta')}-1 \right),
\end{eqnarray}
}

where

\framethis{
\begin{equation}
J_{F,\alpha}(\theta:\theta') = \alpha F(\theta) + (1-\alpha) F(\theta') - F(\alpha\theta+(1-\alpha)\theta')   
\end{equation}
}
is the skew divergence based on the Jensen gap obtained from the log-normalizer convex function $F$. 
$J_{F,\alpha}(\theta:\theta')$ is non-negative for $\alpha\in [0,1]$ and non-positive for $\alpha\in (-\infty,0]\cup [1,\infty)$.
It looses discriminatory power (i.e., $J_F(\theta:\theta')=0,\ \forall\theta,\theta'$) for $\alpha\in\{0,1\}$.

\begin{proof}
Let us consider computing $I_{\alpha}(p:q)=I_{\alpha}(\theta:\theta')$ for members $p\sim E_F(\theta)$ and $q\sim E_F(\theta')$ of the same exponential family $E_F$:

\begin{eqnarray}
I_{\alpha}(p:q) & =&  \int p(x)^\alpha q(x)^{1-\alpha} \dx \\
I_{\alpha}(\theta:\theta') & =& \int \exp^{\alpha(\inner{t(x)}{\theta}-F(\theta)+k(x))}\\
&& \times 
\exp^{(1-\alpha)(\inner{t(x)}{\theta '}-F(\theta')+k(x))} \dx\\
&=&   \int e^{ \inner{t(x)}{\alpha\theta+(1-\alpha)\theta'}} \\
&&\times \exp{-\alpha F(\theta)-(1-\alpha)F(\theta')+k(x)  }\dx\\
&=&  \int e^{F(\alpha\theta+(1-\alpha)\theta')-\alpha F(\theta)}\\
&& \times\exp{-(1-\alpha)F(\theta')}p_F(x;\alpha\theta+(1-\alpha)\theta')\dx \\
&=&  e^{-J_{F,\alpha}(\theta:\theta')}  \underbrace{\int p_F(x;\alpha\theta+(1-\alpha)\theta')\dx}_{=1} \\
&=& e^{-J_{F,\alpha}(\theta:\theta')}  >0
\end{eqnarray}

Thus the R\'enyi divergence of members of the same exponential family amounts to compute a scaled skew Jensen divergence for the log-normalizer:

\begin{equation}
D_\alpha^R(p:q)= \frac{J_{F,\alpha}(\theta:\theta')}{1-\alpha}
\end{equation}
Note that for $\alpha>1$, we have both $1-\alpha<0$ and $J_{F,\alpha}(\theta:\theta')<0$ so that R\'enyi divergence is non-negative.
(However, for $\alpha<0$, $1-\alpha>0$ but  $J_{F,\alpha(\theta:\theta')}<0$. This shows that R\'enyi divergences are defined for $\alpha\in (0,\infty)\backslash\{1\}$.)

The formula for Tsallis divergence follows from the following conversion formula
\begin{equation}
D_\alpha^T(p:q)= \frac{e^{(\alpha-1)D_\alpha^R(p:q)}-1}{\alpha-1}  = \frac{e^{-J_{F,\alpha}(\theta:\theta')}-1}{\alpha-1}
\end{equation}

\end{proof}

Observe that $D_\alpha^T(p:q) \simeq_{\alpha\rightarrow 1} D_\alpha^R(p:q) \simeq_{\alpha\rightarrow 1} \KL(p:q)$ (using the argument $e^{x}\simeq_{x\rightarrow 0} 1+x$).
When $\alpha\rightarrow 1$, we get the well-known result~\cite{RaoNayak:1985,bregmankmeans-2005} that

\framethis{
\begin{equation}
\KL(  p_F(x;\theta) : p_F(x;\theta') ) = B_F(\theta' : \theta),
\end{equation}
}
where $B_F$ is the Bregman divergence~\cite{Bregman67} defined by

\framethis{
\begin{equation}
B_F(p : q)=F(p)-F(q)-\innerproduct{p-q}{\nabla F(q)}.
\end{equation}
}

\begin{proof}
Consider the limit case of R\'enyi divergence of members of the same exponential family as $\alpha\rightarrow 1$.
We shall use the following Taylor expansion (G\^ateaux derivative) of a skew Jensen divergence:

\begin{eqnarray}
J_{F,\alpha}(\theta:\theta') &=& F(\alpha \theta+ (1-\alpha)\theta')\\
&& -\alpha F(\theta)-(1-\alpha)F(\theta') \\
 & \simeq_{\alpha\rightarrow 1}&
(1-\alpha)F(\theta')+(1-\alpha)\inner{\theta'-\theta}{\nabla F(\theta)}\\
&& -(1-\alpha)F(\theta) 
\end{eqnarray}

\begin{eqnarray}
\lim_{\alpha\rightarrow 1} D_\alpha^R(p:q) &= & \lim_{\alpha\rightarrow 1} \frac{J_{F,\alpha}(\theta:\theta')}{1-\alpha} = \KL(p:q)\\
 & \simeq_{\alpha\rightarrow 1} &  F(\theta') - F(\theta)  - \inner{\theta'-\theta}{\nabla F(\theta)} \\
 & = &  B_F(\theta' : \theta)
\end{eqnarray}

\end{proof}

A direct alternative proof is also given in Appendix~\ref{sec:dp}.

\begin{example}
Consider the exponential distribution.
We recall that the natural parameter is $\theta=-\lambda$ and the log-normalizer $F(\theta)=-\log-\theta=-\log\lambda$.
For two members $p\sim E_F(\theta)$ and $q\sim E_F(\theta')$ of the same family of exponential distributions, 
we have the R\'enyi divergence $D^R_\alpha(p:q)=\frac{1}{1-\alpha} (\alpha F(\theta)+(1-\alpha) F(\theta') -F(\alpha\theta+(1-\alpha)\theta'))
= \frac{1}{1-\alpha}\log \frac{\alpha\lambda+(1-\alpha)\lambda'}{\lambda^\alpha \lambda'^{1-\alpha}}$.
The Tsallis divergence is $D^T_\alpha(p:q)=\frac{1}{1-\alpha}(\frac{\lambda^\alpha \lambda'^{1-\alpha}}{\alpha\lambda+(1-\alpha)\lambda'}-1)$.
\end{example}

%%%%%%%%%%
\section{Shannon entropy and cross-entropy for the exponential families}
%%%%%%%%%%
Let us now prove that the Shannon entropy and cross-entropy of distributions belonging to the same exponential family can be expressed as

\framethis{
\begin{eqnarray}
H(p)& = & F(\theta)-\innerproduct{\theta}{\nabla F(\theta)}-E_\theta[k(x)] \\
H(p:q)& = & F(\theta')-\innerproduct{\theta'}{\nabla F(\theta)}-E_\theta[k(x)]
\end{eqnarray}
}

\begin{proof}
Write the relative entropy as the difference of the cross-entropy minus the entropy: 

\begin{equation}
\KL(p : q) = H^\times(p : q)-H(p).
\end{equation}

For distributions belonging to the same exponential families, we can separate the terms independent of $q$ (i.e., $\theta'$) from the terms depending on $p$ (i.e., $\theta$), to get

\begin{eqnarray*}
\KL(p : q) &=& B_F(\theta' : \theta)\\
 & = & F(\theta')-F(\theta)-\innerproduct{\theta'-\theta}{\nabla F(\theta)} \\
&=& \underbrace{F(\theta')-\innerproduct{\theta'}{\nabla F(\theta)}}_{\sim H_F^\times(\theta : \theta')} - \underbrace{( F(\theta)-\innerproduct{\theta}{\nabla F(\theta)} )}_{\sim H_F(\theta)}
\end{eqnarray*}

Since the Bregman convex generator $F$ is defined up to an affine term $ax+b$ in the Bregman divergence, and since the factor $a$ leaves independent both the entropy and cross-entropy terms, we deduce that

\begin{equation}
H(p)=H_F(\theta)=F(\theta)-\innerproduct{\theta}{\nabla F(\theta)}+b,
\end{equation}
where $b$ is a constant.
To determine explicitly the entropic normalization additive constant $b$, we proceed as follows:

\begin{eqnarray}
&&H_F(\theta)  =  -\int p_F(x;\theta) \log p_F(x;\theta) \dx \\
& = & F(\theta)-\int p_F(x;\theta)\innerproduct{t(x)}{\theta}\dx - \int k(x) p_F(x;\theta)\dx\\
& = & F(\theta)-  \Innerproduct{\int p_F(x;\theta) t(x) \dx}{\theta}\dx - \int k(x) p_F(x;\theta)\dx\\
& = & H_F(\theta)=F(\theta)-\innerproduct{\theta}{\nabla F(\theta)} + b
\end{eqnarray}

That is, the constant is given by $b=-\int k(x) p_F(x;\theta)\dx  =  -E_\theta[k(x)]$.
(It depends on the member $\theta$ of the family for $k(x)\not =0$.)
 
\end{proof} 

%\begin{example}
%Consider the Rayleigh distribution
%$p(x;\sigma^2) = \frac{x}{\sigma^2} \exp \left( -\frac{x^2}{2\sigma^2} \right) $
%that belongs to the exponential families for the log-normalizer $F(\theta)=- \log (-2 \theta)$, natural parameter $\theta=-\frac{1}{2\sigma^2}$, sufficient statistic $t(x)=x^2$, derivative (gradient) $F'(\theta) = -\frac{1}{\theta}$ and carrier measure $k(x)=\log x$.
%Let $X$ be a random variable following the Rayleigh distribution.
%We have $H(X)=1+ \ln \frac{\sigma}{\sqrt{2}}+\frac{\gamma}{2}$
%where $\gamma=0.57721566...$ stands for the Euler-Mascheroni constant.
%This is the term related to the carrier measure $\log x$ integrated over the distribution
%\end{example}

\begin{example}
Consider the Poisson distribution with probability mass function 
$p( x ; \lambda ) = \frac{\lambda^x \exp(-\lambda)}{x!}$.
The canonical decomposition yields $\theta=\log\lambda$, $F(\theta)=\exp\theta=\lambda$ (derivative is $F'(\theta)=\exp\theta=\lambda$), $t(x)=x$ and
$k(x)=-\log x!$.
The Poisson entropy is therefore $F(\theta)-\theta F'(\theta) + b = \lambda (1-\log \lambda)-E[k(x)]$.
Since $k(x)=-\log x!$, we have $b=-E[k(x)]=\sum_{k=0}^\infty p_F(x;\lambda)\log k!=e^{-\lambda} \sum \frac{\lambda^k\log k!}{k!}$.
\end{example}

%
%To illustrate those closed-form formula, consider the multivariate Gaussian distribution:
%
%
%This is in accordance with the result derived by analytic integration using a change of variable in~\cite{JensenRenyi-CSPL328-2001}.
%
%Consider now the bi-parametric family of deformed logarithms:
%
%\begin{equation}\label{eq:dl}
%L_{\kappa,r}(x) = x^{r}\frac{x^{\kappa}-x^{-\kappa}}{2\kappa} = -L_{\kappa,-r}(1/x).
%\end{equation}
%
%This framework allows one to conveniently unify Shannon, R\'enyi, Tsallis, Abe, Kaniadakis, and Mittal-Sharma-Taneja entropies~\cite{Kaniadakis-2004} by defining a logarithm-deformed cross-entropy and entropy as
%
%\begin{equation}
%H^K_{\kappa,r}(p,q) &=& - \int p(x) L_{\kappa,r}(q(x)})\dx  \\
%H^K_{\kappa,r}(p) &=& - \int p(x) L_{\kappa,r}(p(x)})\dx  
%\end{equation}
%Two kinds of Tsallis divergence~\cite{Tsallis2Div:2009}.

%%%%%%%%%%%%%%%%%%%%%%%%%%%%
\section{Summary, conclusion and discussion}\label{sec:concl}
%%%%%%%%%%%%%%%%%%%%%%%%%%%%

In this paper, we have given closed-form expressions for the R\'enyi and Tsallis divergences of distributions $p\sim E_F(\theta)$ and $q\sim E_F(\theta')$ belonging to the same exponential family $E_F$:

\begin{eqnarray}
D_\alpha^R(p:q) &=& \frac{J_{F,\alpha}(\theta:\theta')}{1-\alpha}, \\
D_\alpha^T(p:q) & = &  \frac{e^{-J_{F,\alpha}}(\theta:\theta')-1}{\alpha-1}, \\
\KL(p:q) & = & \lim_{\alpha\rightarrow 1} D_\alpha^R(p:q)\\
& = &  \lim_{\alpha\rightarrow 1} D_\alpha^T(p:q)= B_F(\theta' : \theta),
\end{eqnarray}
where 

\begin{eqnarray}
J_{F,\alpha}(\theta:\theta')&=& \alpha F(\theta)+(1-\alpha) F(\theta')- F(\alpha\theta+(1-\alpha)\theta')\\
&=& J_{F,1-\alpha}(\theta':\theta)
\end{eqnarray}
 is the skew Jensen divergence. 
Since the R\'enyi divergence for $\alpha=\frac{1}{2}$ is related to the Bhattacharrya coefficient and Hellinger distance, this also yields closed-form expressions for members of the same exponential family:

\begin{eqnarray}
B(p,q) &=& e^{-J_{F,\frac{1}{2}}(\theta,\theta')}, \\
H(p,q) &=& \sqrt{1-e^{-J_{F,\frac{1}{2}}(\theta,\theta')}}.
\end{eqnarray}

Furthermore, we showed that the R\'enyi and Tsallis entropies, including Shannon entropy in the limit case, can be expressed respectively as 

\begin{eqnarray}
&&H^R_\alpha(p_F(x;\theta)) =  \frac{1}{1-\alpha} \left(F(\alpha\theta)-\alpha F(\theta)+\log E_p[e^{(\alpha-1)k(x)}]\right)   \\
&&H^T_\alpha(p_F(x;\theta)) =   \frac{1}{1-\alpha} \left((e^{F(\alpha\theta)-\alpha F(\theta)}) E_p[e^{(\alpha-1)k(x)}]-1  \right) \\
&&H(p_F(x;\theta))  =   F(\theta)-\inner{\theta}{\nabla F(\theta)}-E_p[k(x)]
\end{eqnarray}
The Shannon cross-entropy is given by

\begin{equation}
H^\times(p_F(x;\theta) :p_F(x;\theta') ) = F(\theta')-\inner{\theta'}{\nabla F(\theta)}  -E_p[k(x)]
\end{equation}

Thus these entropies admit closed-form formula  whenever the normalizing carrier measure is zero ($k(x)=0$):
\begin{eqnarray}
H^R_\alpha(p_F(x;\theta)) &=&  \frac{1}{1-\alpha} \left(F(\alpha\theta)-\alpha F(\theta) \right)   \\
H^T_\alpha(p_F(x;\theta)) &=&   \frac{1}{1-\alpha} \left(e^{F(\alpha\theta)-\alpha F(\theta)} -1  \right) \\
H(p_F(x;\theta))  &= &  F(\theta)-\inner{\theta}{\nabla F(\theta)}
\end{eqnarray}

This includes the case of Bernoulli, exponential, Gaussian and center Laplacian distributions, among others.
(We report in~\ref{sec:mg} the R\'enyi entropy for multivariate Gaussian distributions using matrix calculus.)

Recently, Poczos and Schneider~\cite{alphadivEstimation:2011} have proposed a technique to estimate the $\alpha$-divergence based 
on the $k$-nearest neighbor graph. Although applicable to any kind of distributions, their method is computationally intensive and limited in practice to small dimensions.
In contrast, we may  estimate the R\'enyi entropy and divergence of distributions belonging to the same exponential family by applying the closed-form
expressions on the estimates of parameter distributions.
This is all the more efficient as the maximum likelihood estimator (MLE) of exponential families for independent and identically distributed (i.i.d.) observations $x_1, ..., x_n$ is also available in closed-form:

\begin{equation}
\hat\theta=(\nabla F)^{-1} \left(\frac{1}{n}\sum_{i=1}^n t(x_i)\right).
\end{equation}
This estimate $\hat\theta$ is termed the {\it observed point} in information geometry~\cite{informationgeometry-2000}.

\appendix{R\'enyi and Tsallis entropies and divergences for multivariate Gaussians}\label{sec:mg}
%%%%%%%%%%%%%%

The probability density of a multivariate Gaussian centered at $\mu$ with covariance matrix $\Sigma$ is given by 

\begin{equation}
p(x;\mu,\Sigma) = \frac{1}{(2\pi)^{\frac{d}{2}}\sqrt{|\Sigma|}} \exp -\frac{(x-\mu)^T \Sigma^{-1}  (x-\mu) }{2}
\end{equation}

Let us rewrite this density to fit the canonical decomposition of exponential families:

\begin{eqnarray}
 p(x;\mu,\Sigma) &=& \exp \left(
-\frac{1}{2} x^T \Sigma^{-1} x + x \mu^T\Sigma^{-1} - \frac{1}{2}\mu^T\Sigma^{-1}\mu \right.\\
&& \left. - \frac{1}{2}\log (2\pi)^{d} |\Sigma| 
\right)\nonumber\\
&= & \exp \left( \Inner{(x,x^T x)}{\left(\Sigma^{-1}\mu,-\frac{1}{2}\Sigma^{-1}\right)} - F(\theta) 
\right)
\end{eqnarray}
with $\theta=(\Sigma^{-1}\mu,-\frac{1}{2}\Sigma^{-1})$ and $F(\theta)=\frac{1}{2}\log (2\pi)^d|\Sigma| + \frac{1}{2}\mu^T \Sigma^{-1} \mu$  (and $k(x)=0$).
Natural parameter $\theta=(\Sigma^{-1}\mu, -\frac{1}{2}\Sigma^{-1})=(v,M)$ consists in two parts: a vectorial part $v$, and a symmetric negative definite matrix part $M\preceq 0$.
The inner product of $\theta=(v,M)$  and $\theta'=(v',M')$  is defined as 

\begin{equation}
\inner{\theta}{\theta'} =  v^T v' + \mathrm{tr}(M^T M'),
\end{equation}
where tr denote the matrix trace (i.e., sum of diagonal elements).

Since $|\Sigma|=-\frac{1}{2|M|}$ ($|M|=-\frac{1}{2}|\Sigma^{-1}|=-\frac{1}{2|\Sigma|}$) and $\mu=\Sigma v=-\frac{1}{2}M^{-1}v$,  it follows that the log-normalizer expressed using the natural parameters  is 

\begin{eqnarray}
 F(\mu,\Sigma) & = & \frac{1}{2}\log (2\pi)^d|\Sigma| + \frac{1}{2}\mu^T \Sigma^{-1} \mu  \\
 F(v,M) &=& \frac{1}{2}\log (2\pi)^d+ \frac{1}{2}\log \left(-\frac{1}{2 |M|}\right)  - \frac{1}{4} v M^{-1} v^T
\end{eqnarray}

%Note that for univariate Gaussians, we have $\theta=(\frac{\mu}{\sigma^2},-\frac{1}{2\sigma^2})$, and the log-normalizer parameterized by two scalars as
%\begin{equation}
%F(s_1,s_2) = 
%\end{equation}

Let us now write the term $F(\alpha\theta)$:

\begin{eqnarray}
F(\alpha\theta) &=&  F(\alpha v,\alpha M)\\
&=& \frac{1}{2}\log (2\pi)^d+ \frac{1}{2}\log \left(-\frac{1}{2 |\alpha M|}\right)  - \frac{1}{4}\alpha v^T M^{-1} v
\end{eqnarray}
We shall use the fact that $|\alpha M|=\alpha^d |M|$ for $d$-dimensional matrices.
It follows\footnote{Note that the terms $-\frac{1}{4}(\alpha v)^T (\alpha M)^{-1} (\alpha v) + \frac{1}{4} \alpha v^T M^{-1} v$ vanishes so that R\'enyi entropy does not depend on the mean parameter $\mu$.} that

\begin{eqnarray}
&&H^R_\alpha(\theta) = \frac{1}{1-\alpha} (F(\alpha\theta) - \alpha F(\theta)) \\
&=& \frac{1}{1-\alpha} \left(\frac{d}{2}(1-\alpha)\log 2\pi + \frac{1}{2}\log -\frac{1}{2|\alpha M|} \right.\\
&& \left. -\frac{\alpha}{2}\log -\frac{1}{2|M|}   \right)\\
& = & \frac{d}{2}\log 2\pi + \frac{1}{1-\alpha}  \left( \frac{1}{2}\log|\Sigma|-\frac{d}{2}\log\alpha - \frac{\alpha}{2}\log |\Sigma| \right)\\
& = &  \frac{d}{2}\log 2\pi + \frac{1}{2}\log |\Sigma| - \frac{d\log \alpha}{2(1-\alpha)}
\end{eqnarray}

\appendix{Kullback-Leibler divergence of exponential families  as Bregman divergences}\label{sec:dp}
%%%%%%%%%%%%%%%%%%%%%%%%%%%%

Let us prove that for two distributions $p\sim E_F(\theta)$ and $q\sim E_F(\theta')$ belonging to the same exponential family $E_F$, we have

\begin{equation}
\KL(p:q) = B_F(\theta':\theta)
\end{equation}

\begin{proof}
We first show that $\nabla F(\theta)=E[t(X)]$ with $t(X)$ the sufficient statistics:

\begin{eqnarray}
F(\theta) & =& \log \int_x \exp (\innerproduct{t(x)}{\theta}+k(x)) \dx\\
\nabla F(\theta) &=& \left [
\frac{ \int_x t(x)\exp (\innerproduct{t(x)}{\theta}+k(x) ) \dx }{ \int_x \exp\{\innerproduct{t(x)}{\theta}+k(x)\} \dx }
\right ]_j
\end{eqnarray}
Since $e^{F(\theta)}=\int_x \exp (\innerproduct{t(x)}{\theta}+k(x)) \dx$, we replace the denominator to get
\begin{eqnarray}
\nabla F(\theta) &=& \int_x t(x)\exp\{\innerproduct{t(x)}{\theta}-F(\theta)+k(x)\} \dx\\
&= & \int_x t(x) p_F(x;\theta) \dx \\
&=&  E_{\theta}[t(x)]
\end{eqnarray}

We are now ready to prove $\KL(p:q) = B_F(\theta':\theta)$:

\begin{eqnarray}
\KL(p : q) &=& \int_x p_F(x;\theta)\log \frac{p_F(x;\theta)}{p_F(x;\theta')} \dx \\
& = & \int_x p_F(x;\theta) \left(F(\theta')-F(\theta)+\dotproduct{\theta-\theta'}{t(x)}\right) \dx \\
& = & \int_x p_F(x;\theta)    (B_F(\theta' : \theta)+\\
&&\dotproduct{\theta'-\theta}{\nabla F(\theta)}+\dotproduct{\theta-\theta'}{t(x)} ) \dx \\
& = & B_F(\theta' :\theta)+ \int_x p_F(x;\theta) \dotproduct{\theta'-\theta}{\nabla F(\theta)-t(x)}) \dx\\
& = &  B_F(\theta': \theta)  - \int_x p_F(x;\theta) \dotproduct{\theta'-\theta}{t(x)} \dx\\
&& + \dotproduct{\theta'-\theta}{\nabla F(\theta)} \\
&= & B_F(\theta' : \theta)  -  \dotproduct{\theta'-\theta}{\int_x p_F(x;\theta) t(x)\dx}\\
&& +   \dotproduct{\theta'-\theta}{\nabla F(\theta)}\\
&= & B_F(\theta' : \theta) 
\end{eqnarray}
since $\nabla F(\theta)=E[t(X)]$.

%There exists a bijection between regular Bregman divergences and exponential families~\cite{}.

\end{proof}

%%%
%%% Author biographies
%%%

% Picture is FrankNIELSEN.jpg
\begin{IEEEbiography}{Frank Nielsen}
received the BSc (1992) and MSc (1994)
degrees from Ecole Normale Sup\'erieure (ENS Lyon, France).
He prepared  and defended his PhD  on adaptive
computational geometry at INRIA
Sophia-Antipolis (France) in 1996. 
In 1998, he joined Sony Computer Science
Laboratories Inc., Tokyo (Japan) where he is currently senior researcher. 
He became a professor of the CS Dept. of Ecole Polytechnique in 2008.
His current research
interests include information geometry, vision, graphics, learning, and optimization.
He is a senior ACM and senior IEEE member.
\end{IEEEbiography}

\begin{IEEEbiography}{Richard Nock}
received the agronomical engineering
degree from the Ecole Nationale Superieure
Agronomique de Montpellier, France
(1993), the PhD degree in computer science
(1998), and an accreditation to lead research
(HDR, 2002) from the University of Montpellier
II, France. Since 1998, he has been a faculty
member at the Universite Antilles-Guyane in
Guadeloupe and in Martinique, where his
primary research interests include machine
learning, data mining, computational complexity, and image processing.
\end{IEEEbiography}

\end{document}